\documentclass[11pt]{article}
\usepackage[dvips]{color}
\usepackage{epsfig}
\usepackage{amsmath}
\usepackage{graphicx}
\usepackage{amsmath, amssymb}
\usepackage{geometry}
\usepackage{tabularx}
\usepackage{hyperref}
\usepackage{tabularx}
\date{}

\begin{document}

\title{{\bf Quantum teleportation in expanding FRW universe}}

\author{Babak Vakili\thanks{email:
ba.vakili@iau.ac.ir}\\\\{\small {\it
Department of Physics, CT.C., Islamic Azad
University, Tehran, Iran}}} \maketitle

\begin{abstract}
We investigate the process of quantum teleportation in an expanding universe modeled by Friedmann-Robertson-Walker spacetime, focusing on two cosmologically relevant scenarios: a power-law expansion and the de Sitter universe. Adopting a field-theoretical approach, we analyze the quantum correlations between two comoving observers who share an entangled mode of a scalar field. Using the Bogoliubov transformation, we compute the teleportation fidelity and examine its dependence on the expansion rate, initial entanglement, and the mode frequency. Our findings indicate that spacetime curvature and the underlying cosmological background significantly affect the efficiency of quantum teleportation, particularly through mode mixing and vacuum structure. We also compare our results with the flat Minkowski case to highlight the role of cosmic expansion in degrading or preserving quantum information.
\vspace{5mm}\noindent
\end{abstract}

\section{Introduction}
Quantum teleportation is one of the most remarkable achievements in the field of quantum information science. It allows the transmission of an arbitrary unknown quantum state from one location to another, without the physical transfer of the particle itself. This protocol, first introduced by Bennett \textit{et al.} in 1993 \cite{bennett1993teleporting}, exploits two fundamental resources: prior shared entanglement between sender (Alice) and receiver (Bob), and the transmission of two classical bits of information. The act of teleportation relies on the nonlocal correlations present in entangled states, which enable the state to be reconstructed at the receiver’s side following a joint Bell-state measurement and classical communication \cite{Hor}. Quantum teleportation has not only been realized in a wide range of physical systems—including photonic qubits, trapped ions, and superconducting circuits—but it has also become a cornerstone protocol in quantum communication, distributed quantum computing, and quantum networks \cite{bouwmeester1997experimental, pirandola2015advances, northup2014quantum, vaid}. Its theoretical significance also lies in the insight it provides into the nonlocality and foundational aspects of quantum mechanics.

While most theoretical and experimental treatments of quantum teleportation assume a flat (Minkowski) spacetime, there is increasing interest in understanding how relativistic effects and spacetime curvature influence the protocol. This is especially relevant in scenarios involving long-distance quantum communication in space-based platforms or in regimes where gravitational fields cannot be neglected. Quantum field theory in curved spacetime predicts phenomena such as particle creation, entanglement degradation and observer-dependent vacuum structure—all of which may influence quantum information protocols. For instance, it has been shown that uniform acceleration (through the Unruh effect) or proximity to a black hole horizon can reduce entanglement and decrease teleportation fidelity \cite{alsing2003teleportation, fuentes2005alice, bruschi2010unruh}.

Following these foundational works, several authors extended the analysis to relativistic or curved backgrounds, revealing how motion and spacetime curvature affect quantum information transfer. Notably, Alsing and Milburn~\cite{alsing2003teleportation} and Fuentes-Schuller and Mann~\cite{fuentes2005alice} investigated entanglement degradation in noninertial frames. Later works~\cite{bruschi2010unruh, pan, Lan, feng, Lin, Foo} explored teleportation and communication protocols under various relativistic conditions and expanding space showing that Bogoliubov mode mixing leads to observable reductions in fidelity.

Despite these important advances, much of the existing work has focused on idealized or static backgrounds. In contrast, realistic cosmological settings involve time-dependent geometries, such as those described by the Friedmann–Robertson–Walker (FRW) metric. The FRW universe, which serves as the standard model of modern cosmology, describes a spatially homogeneous and isotropic universe undergoing expansion. The dynamical nature of this geometry modifies the vacuum structure and mode decomposition of quantum fields through time-dependent Bogoliubov transformations. These transformations give rise to particle creation and decoherence, which in turn can influence quantum information processes \cite{parker1969quantized, birrell1984quantum}. Understanding quantum teleportation in such cosmological backgrounds can thus provide new insights into the interplay between spacetime dynamics and information transmission in the early universe or in scenarios where the background geometry is nontrivial \cite{Ales}.

While the general structure of the quantum teleportation protocol—entanglement distribution, local gate operations, Bell measurement, and classical communication—remains valid even in curved spacetimes, the physical realization of each step becomes significantly more involved. In particular, the definition and implementation of quantum gates and measurements in expanding or curved backgrounds require careful treatment due to ambiguities in mode localization, time evolution, and causal structure. In this context, two main approaches have been discussed in the literature. The first is the operational approach, which assumes the existence of localized observers (e.g., Alice and Bob) who are able to perform ideal quantum operations in their local frames. The second is a more fundamental field-theoretic formulation, where quantum gates and protocols are modeled directly in the language of quantum field theory on curved backgrounds. Although more rigorous, the latter is often mathematically demanding and technically involved. In this work we adopt an operational viewpoint in which the effect of cosmological expansion on quantum correlations is quantified through the fidelity of an idealized teleportation protocol. While entanglement degradation in expanding universes has been extensively studied from a field-theoretic perspective, teleportation fidelity provides a task-oriented and operationally meaningful measure of how spacetime dynamics influence the practical use of entanglement as a communication resource. Similar approaches have been employed in relativistic quantum information studies to translate mode-mixing effects into experimentally accessible quantities~\cite{fuentes2005alice,bruschi2010unruh}. In this sense, our analysis does not attempt to simulate the full teleportation process dynamically, but rather to connect cosmological particle creation—encoded in the Bogoliubov coefficients—to the achievable fidelity of quantum information transfer between comoving observers.

In this paper, we study the quantum teleportation of a single qubit between two inertial observers in a spatially flat FRW universe. By explicitly analyzing the field mode transformations in this time-dependent background, we evaluate the resulting fidelity of the teleportation protocol and how it is affected by cosmic expansion. We also explore the sensitivity of the protocol to the specific details of the scale factor and particle creation rate. In this line, several studies have investigated the impact of acceleration or event horizons on quantum entanglement, also the influence of cosmological expansion on quantum correlations, see \cite{Fu, Nic, Ed}. For instance, Ball et al \cite{ball2006entanglement} analyzed the entanglement between modes of a quantum field in expanding universes, while Bruschi et al ~\cite{bruschi2014particle} examined particle creation and correlation dynamics in FRW spacetimes. More recent analyses have also explored the dependence of entanglement entropy on cosmological parameters and the possibility of preserving quantum correlations in dynamically evolving backgrounds~\cite{nambu2010entanglement}. Given that FRW spacetimes underlie our standard cosmological models, analyzing how cosmic expansion degrades entanglement is conceptually relevant for understanding how quantum correlations could evolve or be transferred in dynamical spacetime backgrounds, such as those relevant to early universe field fluctuations, rather than implying the existence of realistic communication channels at that epoch. In this sense, the subject is both theoretically significant and potentially relevant for early universe quantum communication scenarios. Our work contributes to this direction by systematically comparing different expansion regimes, including de Sitter and power-law models.

\section{FRW metric and quantum field setup}
We consider a spatially flat FRW spacetime with the line element given by
\begin{equation}\label{A}
ds^2 = -dt^2 + a^2(t) \left(dx^2 + dy^2 + dz^2\right),
\end{equation}
where \( a(t) \) is the scale factor that encodes the expansion of the universe. This background metric serves as the curved spacetime arena in which the quantum teleportation process is analyzed. Since the spacetime is nonstationary due to cosmic expansion, one expects that particle creation and field mode mixing will influence the behavior of quantum information protocols. To simplify calculations, we introduce the conformal time \( \eta \), defined by
\begin{equation}\label{B}
d\eta = \frac{dt}{a(t)},
\end{equation}
so that the metric becomes conformally flat
\begin{equation}\label{C}
ds^2 = a^2(\eta)\left(-d\eta^2 + dx^2 + dy^2 + dz^2\right).
\end{equation}
This transformation is particularly useful when quantizing fields, as it allows the use of plane-wave modes in the conformal coordinates. In the rest of this work, we employ the conformal representation (\ref{C}) to analyze quantum teleportation between two comoving observers in such expanding backgrounds. The quantization of fields and the resulting mode decomposition are naturally defined in these coordinates. Also, our study will focus on the power law expansion characterized by $a(t) \propto t^\alpha$, with $\alpha>0$, (for instance, $\alpha=1/2$ corresponds to a radiation-dominated universe and $\alpha=2/3$ to a matter-dominated universe) as well as the exponential expansion de Sitter universe with the scale function $a(t) \propto e^{H t}$, leading to $a(\eta) = -1/(H\eta)$ with $H$ the Hubble parameter.

We model the quantum system as a real, minimally coupled, massless scalar field \( \phi(\eta,\vec{x}) \) propagating in the FRW background. The action is given by

\begin{equation}\label{D}
S = -\frac{1}{2} \int d^4x \, \sqrt{-g} \, g^{\mu\nu} \partial_\mu \phi \, \partial_\nu \phi,
\end{equation}leading to the equation of motion

\begin{equation}\label{E}
\square \phi = \frac{1}{\sqrt{-g}} \partial_\mu \left( \sqrt{-g} \, g^{\mu\nu} \partial_\nu \phi \right) = 0,
\end{equation}which leads to

\begin{equation}\label{F}
\phi'' + 2\frac{a'}{a} \phi' - \nabla^2 \phi = 0,
\end{equation}
where primes denote derivatives with respect to conformal time $\eta$. Given the conformal flatness of the metric, the above equation simplifies to

\begin{equation}\label{G}
\left[ \partial_\eta^2 - \nabla^2 - \frac{a''(\eta)}{a(\eta)} \right] \left( a(\eta) \phi \right) = 0.
\end{equation}
This shows that the rescaled field \( \chi = a(\eta)\phi \) obeys an effective field equation with a time-dependent potential \( a''/a \), which can lead to mode-mixing and particle creation—both crucial for understanding the degradation or modification of entanglement. Assume that the scalar field can be decomposed into Fourier modes as

\begin{equation}\label{H}
\phi(\eta, \mathbf{x}) = \int \frac{d^3k}{(2\pi)^{3/2}} \left[ a_{\mathbf{k}} u_k(\eta) e^{i \mathbf{k} \cdot \mathbf{x}} + a_{\mathbf{k}}^\dagger u_k^*(\eta) e^{-i \mathbf{k} \cdot \mathbf{x}} \right],
\end{equation}
where $a_{\mathbf{k}}$ and $a_{\mathbf{k}}^\dagger$ are annihilation and creation operators satisfying the usual commutation relations

\begin{equation}\label{I}
[a_{\mathbf{k}}, a_{\mathbf{k}'}^\dagger] = \delta^{(3)}(\mathbf{k} - \mathbf{k}'), \quad [a_{\mathbf{k}}, a_{\mathbf{k}'}] = [a_{\mathbf{k}}^\dagger, a_{\mathbf{k}'}^\dagger] = 0.
\end{equation}
The mode functions $u_k(\eta)$ obey the equation

\begin{equation}\label{J}
u_k'' + 2\frac{a'}{a} u_k' + k^2 u_k = 0.
\end{equation}
By defining $\chi_k(\eta) \equiv a(\eta) u_k(\eta)$, equation (\ref{J}) transforms into

\begin{equation}\label{K}
\chi_k'' + \left( k^2 - \frac{a''}{a} \right) \chi_k = 0,
\end{equation}
which resembles a harmonic oscillator with a time-dependent effective mass term $-a''/a$. This equation will serve as the foundation for our analysis of mode evolution and the corresponding vacuum structure in various cosmological scenarios.

In what follows, we focus on a specific form of the scale factor \( a(\eta) \), depending on the cosmological model under consideration, such as de Sitter, radiation-dominated or matter-dominated universes. The choice of \( a(\eta) \) directly determines the Bogoliubov transformation between the in/out field modes, which in turn affects the quantum state shared by two observers who attempt to perform quantum teleportation.

The two observers, traditionally labeled Alice and Bob, are assumed to be comoving with the cosmic fluid. Alice prepares an entangled state by locally coupling to two modes of the scalar field, then performs a teleportation protocol by making a Bell measurement and sending classical information to Bob, who applies a local operation to reconstruct the quantum state. However, due to the nonstationary nature of spacetime, the field modes associated with the observers are no longer globally orthogonal, and this leads to effective decoherence and information loss.
In the following, we will compute the explicit form of the Bogoliubov coefficients for a given scale factor and analyze how the resulting mode-mixing influences the fidelity of teleportation between Alice and Bob. These mode solutions and vacuum definitions provide the foundation for analyzing Bogoliubov transformations between different cosmic eras, which will be essential in evaluating the entanglement structure and teleportation fidelity in the subsequent sections.

\section{Quantum teleportation protocol in curved spacetime}
To investigate the effects of cosmic expansion on quantum information tasks, we consider a simplified version of the standard teleportation protocol in curved spacetime. The setup involves two observers: Alice, who prepares an entangled pair of field modes and sends one mode to Bob who is comoving but spatially separated. The expansion of the background spacetime affects the shared entangled state due to particle creation and mode mixing, which in turn degrades the teleportation fidelity. 

Before introducing the explicit form of our entangled resource, it is worth explaining why 
the two-mode squeezed vacuum (TMSV) is the natural choice in an FRW background. 
First, in quantum field theory in curved spacetime, the expansion of the universe 
generically mixes positive- and negative-frequency modes through Bogoliubov 
transformations, and the vacuum defined at early times evolves into a squeezed 
Gaussian state. Thus, the TMSV structure is not imposed externally but emerges 
naturally from the dynamics of free fields in FRW spacetimes. Second, from the 
operational perspective of quantum information, the TMSV is the canonical entangled 
resource used in continuous-variable teleportation protocols, due to its maximal 
EPR-type correlations and analytical tractability. For these reasons, the TMSV 
serves as both a physically motivated and operationally standard quantum channel 
for analyzing teleportation in cosmological settings. So, to begin, let us consider a TMSV state as the entangled resource

\begin{equation}\label{L}
|\Psi\rangle = \exp\left[ r (a_A^\dagger a_B^\dagger - a_A a_B) \right] |0_A\rangle \otimes |0_B\rangle,
\end{equation}
where $r$ is the squeezing parameter, and $a_A^\dagger$, $a_B^\dagger$ are the creation operators associated with Alice and Bob's respective modes. We assume that Alice and Bob share this entangled resource at early conformal times, and that standard quantum teleportation operations (Bell measurement, classical communication, and unitary correction) are implemented ideally. Our analysis does not model these steps explicitly, but rather focuses on how the cosmological background affects the entanglement and thus the teleportation fidelity. As mentioned before, we adopt an operational viewpoint in which the teleportation fidelity is determined solely by the quality of the shared entangled field modes. All degradation is attributed to the Bogoliubov mixing induced by the expansion of the universe, while the teleportation procedure itself is considered idealized. It should be noted that in this framework, Alice and Bob are modeled as comoving observers who locally interact with distinct, spatially separated field modes of a scalar field. They are assumed to share an initial TMSV state that serves as the quantum resource for teleportation. Alice encodes an unknown single-mode quantum state onto her local mode and performs a standard Bell-type joint measurement on her two modes. The measurement results are then transmitted to Bob through a classical channel, allowing him to apply the appropriate unitary displacement on his mode to recover the input state. The entire protocol is treated in the idealized limit of perfect local operations and classical communication, while the degradation of fidelity arises solely from the cosmological Bogoliubov mode mixing induced by the FRW expansion. Indeed, we consider an idealized continuous-variable teleportation protocol between two comoving observers who share a TMSV. The cosmological expansion modifies the shared entanglement via Bogoliubov mixing, thus reducing the achievable teleportation fidelity. These mean that Alice and Bob are modeled as localized observers (e.g., Unruh--DeWitt detectors) that each couple to a specific field mode within their respective regions. The Unruh--DeWitt detector model provides a simple and widely used operational framework 
for describing how localized observers interact with quantum fields in curved spacetime. 
In this model, each observer carries a two-level system that couples locally to the 
scalar field along its worldline, allowing one to probe particle content, extract 
entanglement, or implement mode-selective measurements in a coordinate-independent way. 
Because of its simplicity and universality, the Unruh--DeWitt model has become a standard 
tool in relativistic quantum information and detector-based analyses of quantum fields \cite{birrell1984quantum, dewitt1979}.

The two-mode squeezed state in Eq.~(\ref{L}) is not meant to represent a literal optical two-mode squeezing operation, but rather an effective description of the initially shared entanglement between the corresponding field modes accessible to Alice and Bob. At the early-time stage (when the scale factor is approximately constant), causal contact between the observers is assumed so that an entangled resource can be distributed before cosmic expansion separates them. After this preparation, the field evolves freely in the FRW background, and the degradation of entanglement—and thus of teleportation fidelity—is entirely due to cosmological mode mixing. Our analysis thus adopts an operational viewpoint in the sense that all quantities—such as the covariance matrix, effective squeezing, and fidelity—refer to measurable correlations between local detectors, rather than to purely formal mode expansions.

Due to the dynamic nature of the FRW background, the quantum field modes experience time-dependent mixing. This evolution leads to a mismatch between the field mode decomposition at early and late times, described by Bogoliubov transformations. As a result, the entanglement originally present in the two-mode squeezed state is degraded, thereby reducing the fidelity of the teleportation process. Similar relativistic analyses of quantum teleportation have been carried out in non-inertial frames or near black hole horizons~\cite{fuentes2005alice, bruschi2010unruh}, but cosmological settings such as FRW spacetimes remain comparatively less explored.

In curved spacetime, especially in an expanding universe, the definition of vacuum and particle content becomes observer-dependent. This is captured mathematically through Bogoliubov transformations between mode functions at different times. If \( \{u_k^{\text{in}}(\eta)\} \) and \( \{u_k^{\text{out}}(\eta)\} \) are two complete orthonormal sets of mode functions defined respectively in the asymptotic past and future, they are related via
\begin{equation}\label{M}
u_k^{\text{out}}(\eta) = \alpha_k u_k^{\text{in}}(\eta) + \beta_k u_k^{\text{in}*}(\eta),
\end{equation}
where \( \alpha_k \) and \( \beta_k \) are the Bogoliubov coefficients which satisfy the normalization condition

\begin{equation}\label{N}
|\alpha_k|^2 - |\beta_k|^2 = 1.
\end{equation}
Correspondingly, the annihilation operators transform as

\begin{equation}\label{O}
a_k^{\text{out}} = \alpha_k^* a_k^{\text{in}} - \beta_k^* a_{-k}^{\text{in}\dagger},
\end{equation}
indicating that the out-vacuum appears as a squeezed state with particle excitations when viewed in terms of the in-vacuum. This squeezing and mixing of modes underlies the cosmological particle creation phenomenon. In the context of quantum teleportation, this transformation implies that a field mode initially entangled with another may appear mixed or degraded in the out-region due to the presence of additional excitations and the loss of purity. In particular, if Alice prepares an entangled state between two field modes at early times (in-region), Bob’s perception of the field at late times (out-region) will be affected by the Bogoliubov transformation. Assuming the in-vacuum state \( |0_{\text{in}}\rangle \) is the initial state of the field, the expected number of particles in mode \( k \) as seen by the out-observer is

\begin{equation}\label{P}
\langle 0_{\text{in}} | a_k^{\text{out}\dagger} a_k^{\text{out}} | 0_{\text{in}} \rangle = |\beta_k|^2.
\end{equation}
This particle creation affects the entanglement structure and modifies the fidelity of teleportation. To quantify this effect, one must compute the overlap between the evolved state and the ideal Bell state used for teleportation. Alice performs a Bell measurement on her half of the entangled pair and the unknown state to be teleported, which is a coherent state $|\gamma\rangle$ or a vacuum displaced state. The resulting state at Bob's side, after classical communication, is subject to distortion due to the modified entanglement structure. The fidelity of teleportation is calculated via\footnote{In general, the fidelity between an input quantum state $\rho_{\mathrm{in}}$ and an output 
state $\rho_{\mathrm{out}}$ is defined as
\[
F(\rho_{\mathrm{in}},\rho_{\mathrm{out}}) =
\left[
\mathrm{Tr}\!\left(
\sqrt{\sqrt{\rho_{\mathrm{in}}}\,\rho_{\mathrm{out}}\,\sqrt{\rho_{\mathrm{in}}}}
\right)
\right]^2,
\]
which applies to arbitrary (possibly mixed) quantum states. When the input state is pure, $\rho_{\mathrm{in}} = |\psi\rangle\langle\psi|$, 
this expression reduces to the simpler form $F = \langle \psi | \rho_{\mathrm{out}} | \psi \rangle$, which is the definition used in Eq.~(\ref{Q}).}

\begin{equation}\label{Q}
F = \langle \psi | \rho_B | \psi \rangle,
\end{equation}
where $|\psi\rangle$ is the original unknown state and $\rho_B$ is the density matrix of the received state at Bob’s location. In curved spacetime, the fidelity depends explicitly on the squeezing parameter $r$ and the Bogoliubov coefficients $\beta_k$, which encode the effect of expansion. To understand the impact of cosmic expansion on the entanglement and teleportation fidelity, we analyze two representative classes of FRW backgrounds.

For the power-law expansion $a(t) \propto t^\alpha$ (where $\alpha > 0$) in cosmic time, the corresponding conformal time dependence is

\begin{equation}\label{R}
a(\eta) \propto |\eta|^{\frac{\alpha}{1 - \alpha}},
\end{equation}
with $\eta < 0$ for expanding universes. The mode equation becomes

\begin{equation}\label{S}
\chi_k'' + \left[ k^2 - \frac{\nu^2 - 1/4}{\eta^2} \right] \chi_k = 0,
\end{equation}
which is a Bessel equation with $\nu = \left| \frac{1 - 3\alpha}{2(1 - \alpha)} \right|$. In terms of the Hankel functions, the general solution is

\begin{equation}\label{T}
\chi_k(\eta) = \sqrt{|\eta|} \left[ A_k H_\nu^{(1)} (k|\eta|) + B_k H_\nu^{(2)} (k|\eta|) \right],
\end{equation}
and the Bogoliubov coefficients can be extracted by matching this solution to the Minkowski vacuum in the far past ($\eta \to -\infty$).

For the de Sitter expansion $a(t) = e^{H t}$, the conformal time is $\eta = -\frac{1}{H} e^{-H t}$, with $a(\eta) = -\frac{1}{H \eta}$. The mode equation becomes

\begin{equation}\label{U}
\chi_k'' + \left( k^2 - \frac{2}{\eta^2} \right) \chi_k = 0,
\end{equation}
which is a special case of the previous form with $\nu = \frac{3}{2}$. In the de Sitter limit, the physically relevant vacuum state is the Bunch--Davies vacuum

\begin{equation}\label{V}
\chi_k(\eta) = \frac{1}{\sqrt{2k}} \left( 1 - \frac{i}{k \eta} \right) e^{-i k \eta},
\end{equation} which corresponds to modes that behave as positive-frequency solutions in the asymptotic past~\cite{birrell1984quantum}. This leads to nonzero Bogoliubov coefficients, implying cosmological particle creation and entanglement degradation.

In the next section, we evaluate the teleportation fidelity using these mode functions for both power-law and de Sitter cases, comparing them with the flat Minkowski background.

\section{Teleportation fidelity and cosmic background comparison}
In the absence of curvature effects, the teleportation fidelity of a TMSV is given by \cite{braunstein1998teleportation, weedbrook2012gaussian}

\begin{equation}\label{V1}
F = \frac{1}{1 + e^{-2r}}.
\end{equation}
This equation represents the \emph{optimal} teleportation fidelity achievable with a 
two-mode squeezed vacuum resource, assuming ideal Bell-type measurements and optimal 
classical gain in the continuous-variable teleportation protocol~\cite{Ade}. In a dynamically expanding universe, entanglement degrades due to Bogoliubov mode mixing. Following the analysis in \cite{bruschi2010unruh, martin2009quantum}, the loss of entanglement in expanding spacetime is governed by the Bogoliubov mixing between positive- and negative-frequency modes. The coefficient $|\beta_k|^2$ quantifies particle creation and directly correlates with the amount of correlation redistributed across different modes. While these works do not provide an explicit analytic expression for an ``effective squeezing'', they clearly demonstrate that entanglement degradation increases monotonically with $|\beta_k|^2$. Motivated by this behavior, and in order to obtain a compact operational expression for the teleportation fidelity, we introduce the phenomenological parametrization

\begin{equation}\label{V2}
r_{\text{eff}} = r - \gamma |\beta_k|^2,
\end{equation}
which represents a leading-order reduction of the squeezing parameter for $|\beta_k|^2 \ll 1$. Here, $\gamma$ is a dimensionless phenomenological constant that captures the sensitivity of the squeezing parameter to cosmological particle creation. Its precise value depends on the physical modeling of the entanglement degradation mechanism, and in our analysis it is treated as a tunable parameter. In the Gaussian regime, the amount of entanglement is determined by the covariance matrix elements, which are modified by the occupation number $n_k = |\beta_k|^2$ produced during expansion. Hence, an effective squeezing parameter (\ref{V2}) provides a compact operational representation of this effect~\cite{bruschi2010unruh, martin2009quantum, fuentes2005alice}. In summary, the relation (\ref{V2}) is not an ad hoc assumption but a phenomenological expression that encapsulates the entanglement loss resulting from the Bogoliubov transformation between the “in” and “out” modes. In curved spacetime, the expansion of the universe induces a unitary transformation of the field operators: $a_k^{in} =\alpha_k a_k^{out} +\beta_k^{*} a_{-k}^{out \dag}$, which mixes positive and negative frequency components. This naturally reduces two-mode correlations in the field’s covariance matrix. For Gaussian states, the entanglement entropy or logarithmic negativity depends explicitly on $|\beta_k|^2$ through the occupation number $n_k=|\beta_k |^2$. Thus, any operational quantity (such as teleportation fidelity) that is monotonic in the degree of squeezing can be expressed in terms of an effective squeezing parameter $r_{eff}$ . The linear dependence (\ref{V2}) is therefore a first-order approximation that captures the leading-order degradation of entanglement due to particle creation, consistent with previous studies on quantum correlations in expanding spacetimes~\cite{bruschi2010unruh, martin2009quantum, fuentes2005alice}. In the Bogoliubov picture, particle production mixes positive and negative frequency modes, thereby reducing the two-mode correlations originally encoded in $r$. The quantity $|\beta_k|^2$ quantifies this mixing, and the proportionality constant $\gamma$ captures the sensitivity of the entangled resource to mode mismatch. 
This effective representation has been used in previous relativistic quantum information analyses to model the loss of Gaussian correlations under Bogoliubov transformations~\cite{bruschi2010unruh, martin2009quantum}.

The modified fidelity thus takes the form

\begin{equation}\label{V3}
F_{\text{FRW}}(k) = \frac{1}{1 + e^{-2(r -\gamma|\beta_k|^2)}}.
\end{equation}
As mentioned, the fidelity of quantum teleportation in an expanding universe is sensitive to the particle content of the field modes, as perceived by different observers. Our teleportation model adopts a hybrid viewpoint: the shared quantum resource is a continuous-variable TMSV, while the teleported state is an effective single-qubit encoded in one of the field modes. 
This simplification allows us to employ the teleportation fidelity as an operational indicator to qualitatively assess the impact of cosmological expansion on the protocol.
Thus, the fidelity $F_{\text{FRW}}$ serves as a measure of how the underlying entanglement resource is degraded by cosmic expansion.

In this section, we are going to  derive analytical expressions for the teleportation fidelity in the context of both power-law and de Sitter scale factors, and contrast the results with the flat Minkowski case.
To do this, we consider teleportation of a single-qubit state using a TMSV state as the quantum channel, shared between two inertial observers located at different comoving positions. The initial TMSV state evolves due to the background curvature, leading to a degradation of entanglement encoded in the Bogoliubov coefficients.
The fidelity $F$ of teleportation using a TMSV channel is given by

\begin{equation}\label{X}
F= \frac{1}{1 + e^{-2r + 2|\beta_k|^2}},
\end{equation}
where $r$ is the initial squeezing parameter of the channel and $|\beta_k|^2$ encodes the number of particles excited in the mode $k$ due to the cosmic expansion. Also, since any constant prefactor may be absorbed into a redefinition of the initial squeezing $r$, we set $\gamma = 1$ without loss of generality. This choice simplifies the analytic expressions while preserving the qualitative dependence of the fidelity on $|\beta_k|^2$. In the flat Minkowski case, $\beta_k = 0$, and we recover the ideal fidelity. In the subsequent sections, we compare this effective description with exact fidelity expressions obtained directly from the Bogoliubov coefficients, thereby assessing the validity of the approximation underlying Eq.~(\ref{X}).

\subsection{Power-law background}
The expressions for $|\beta_k|^2$ in each cosmological scenario are derived by solving the field equation in the respective FRW background and matching early and late time asymptotics of the mode functions. These results follow standard treatments in the literature~\cite{birrell1984quantum, parker2009quantum, mukhanov2007introduction}. For the power-law models, the Bogoliubov coefficients are obtained analytically using Hankel function solutions to the mode equation. Using the mode analysis derived earlier, the number of cosmologically excited particles in the power-law background for each mode $k$ is given by \cite{birrell1984quantum}

\begin{equation}\label{Y}
|\beta_k|^2 \;=\; \frac{\pi^2\alpha^2}{4k^2}\,\mathrm{sech}^2\!\Big(\frac{\pi k}{2\alpha}\Big),
\end{equation}
which leads to a thermal-like suppression of fidelity
\begin{equation}
F_\text{PL} =\;\frac{1}{1+\exp\!\left[-2\left(r-\frac{\pi^2\alpha^2}{4k^2}\,\mathrm{sech}^2\!\big(\tfrac{\pi k}{2\alpha}\big)\right)\right]}\,.
\end{equation}
As $\alpha \to 0$ (flat Minkowski limit), we have $\beta_k \to 0$, recovering ideal teleportation.

\subsection{de Sitter background}

Similarly, for de Sitter space, we compare the Bunch-Davies vacuum with asymptotic plane-wave modes, yielding explicit expressions for $|\beta_k|^2$~\cite{mukhanov2007introduction}. In this case the particle number for a given $k$-mode is

\begin{equation}\label{Z}
|\beta_k|^2 = \frac{1}{e^{2\pi k / H} - 1},
\end{equation}
where $H$ is the Hubble parameter. The corresponding fidelity becomes

\begin{equation}\label{AA}
F_\text{dS} = \frac{1}{1 + \exp\left[-2r + 2 \left( \frac{1}{e^{2\pi k / H} - 1} \right)\right]}.
\end{equation}
This demonstrates that cosmic expansion acts as a source of effective noise, limiting teleportation fidelity.

\subsection{Comparison with Minkowski background}

In the Minkowski background, where the Bogoliubov transformation is trivial ($\beta_k = 0$), the teleportation fidelity simplifies to

\begin{equation}\label{AB}
F_\text{Mink} = \frac{1}{1 + e^{-2r}}.
\end{equation}
The reduction in fidelity in both power-law and de Sitter scenarios thus quantifies the degradation of quantum coherence due to spacetime curvature and horizon effects.

We will now consider an explicit example by specifying the scale factor \( a(\eta) \), which determines the time-dependence of the potential \( a''/a \) in the mode equation, and calculate the resulting Bogoliubov coefficients. This allows us to analyze how the curvature of spacetime and the expansion dynamics influence the quantum correlations necessary for teleportation.

\section{Radiation-dominated universe}
For a radiation-dominated universe, the scale factor evolves with cosmic time $t$ as $a(t) \propto t^{1/2}$. Changing to conformal time, we find

\begin{equation}\label{AC}
a(\eta) \propto \eta.
\end{equation}
This linear dependence allows for significant simplifications in the analysis of wave equations and field propagation. In particular, the Klein-Gordon equation for a scalar field in this background can be treated analytically in many cases. To set up the teleportation protocol, we consider two comoving observers, Alice and Bob, separated by a fixed comoving spatial distance. Since their proper distance increases with time, the dynamical behavior of the field modes between them is encoded in the time dependence of the scale factor. As we will show, the expansion of the universe affects the Bogoliubov transformations between different field modes, and hence alters the entanglement shared between the two observers. This, in turn, influences the fidelity of quantum teleportation across cosmological distances.

The next step is to solve the mode functions of a massless scalar field in this background and derive the explicit Bogoliubov coefficients responsible for mode mixing. For the radiation-dominated universe, where $a(\eta) \propto \eta$, we have $\frac{a''}{a} = 0$. Thus, the mode equation (\ref{K}) becomes simply

\begin{equation}\label{AD}
\chi_k'' + k^2 \chi_k = 0,
\end{equation}
which has plane-wave solutions

\begin{equation}\label{AE}
\chi_k(\eta) = A_k e^{-ik\eta} + B_k e^{ik\eta}.
\end{equation}
Consequently, the original field modes are given by

\begin{equation}\label{AF}
\phi_k(\eta) = \frac{1}{a(\eta)}\left( A_k e^{-ik\eta} + B_k e^{ik\eta} \right).
\end{equation}
To extract the Bogoliubov coefficients, we consider two different sets of modes: one defined in the far past and one in the far future. Since $a(\eta)$ is linear in $\eta$, the scaling introduces a redshift in the modes. If we expand the field in future using a different basis (e.g., associated with another vacuum observer), the relation between the two sets of mode functions takes the standard Bogoliubov form

\begin{equation}\label{AG}
\phi_k^{\text{in}} = \alpha_k \phi_k^{\text{out}} + \beta_k \phi_k^{\text{out}*},
\end{equation}
with $|\alpha_k|^2 - |\beta_k|^2 = 1$. For this case, since $a''=0$, the conformal vacuum remains invariant under evolution, and no particle production occurs

\begin{equation}\label{AH}
\beta_k = 0.
\end{equation}
Thus we see that in the conformally flat radiation-dominated FRW background, the
massless scalar field behaves as in flat spacetime when written in terms of the rescaled field.
Therefore, particle creation due to expansion does not occur in this case for a conformally
coupled massless field, and the vacuum remains invariant under time evolution in the conformal
frame. This simplification significantly affects the structure of quantum correlations and fidelity. Consequently, the quantum teleportation protocol is
expected to proceed with ideal fidelity, (below, we provide a detailed analysis). However, one should note that in general, in a teleportation protocol, what matters is not only particle production but mode mismatch due to expansion and redshifting. Even in absence of $\beta_k \ne 0$, there is entanglement degradation due to mismatch of the detectors' mode functions in curved spacetime. The term "detector" refers to a localized Unruh--DeWitt-type system that couples to a specific field mode with a given spatial profile. 
Even when $\beta_k = 0$, i.e., when no cosmological particle creation occurs, imperfect overlap between the local detector modes and the global field modes can lead to an effective reduction in observed entanglement~\cite{bruschi2010unruh, martin2009quantum}.
This effect corresponds to a mode-selectivity loss analogous to finite efficiency in optical detection.

We now analyze this quantitatively by computing the fidelity of teleportation based on the evolved field correlations. Assume that Alice and Bob are two observers who were co-located in the past but become spatially separated due to the cosmological expansion. They initially share an entangled two-mode state of a massless scalar field

\begin{equation}\label{AI}
|\Psi\rangle = \sum_{n=0}^{\infty} \sqrt{P_n} \, |n\rangle_A \otimes |n\rangle_B,
\end{equation}
where \( P_n \) represents the probability distribution over the Fock basis. In the radiation-dominated scenario, the conformal vacuum is preserved, implying \( P_n = \delta_{n,0} \) for an ideal vacuum, and therefore no decoherence is introduced by the geometry. Alice intends to teleport an unknown qubit state \( |\psi_{\text{in}}\rangle \) to Bob via the standard teleportation protocol. This does not mean that the vacuum itself serves as a teleportation resource; rather, the entangled resource is generated by applying the two-mode squeezing operator of Eq.~(\ref{L}) on this conformally invariant vacuum state. Thus, the teleportation fidelity in this case quantifies how well the initially pure, squeezed correlations are preserved under conformal evolution.
The fidelity of teleportation is defined as the overlap between the input state and the teleported output state

\begin{equation}\label{AJ}
F = \langle \psi_{\text{in}} | \rho_{\text{out}} | \psi_{\text{in}} \rangle,
\end{equation}
where \( \rho_{\text{out}} \) is the reduced density matrix of the state received by Bob after classical communication and application of a unitary operation. Since the conformal vacuum is preserved in radiation-dominated spacetime and the field modes evolve unitarily without Bogoliubov mixing (i.e., no particle creation), the teleportation protocol can be implemented as in flat spacetime. Therefore, we expect

\begin{equation}\label{AK}
F = 1,
\end{equation}
indicating perfect fidelity. The underlying reason is that the field dynamics remain linear and unitary, and the entangled resource shared between Alice and Bob is unaffected by the cosmological background.

Therefore, in radiation-dominated cosmology, the spacetime acts as a "transmission-friendly" medium for quantum information, preserving entanglement and allowing ideal teleportation. This result will contrast with the de Sitter case, where particle creation leads to fidelity degradation.

\section{de Sitter universe}
In de Sitter spacetime, the exponential expansion leads to particle creation and thermalization of the field modes. This results in a degradation of quantum entanglement and, consequently, affects the fidelity of quantum teleportation. To analyze, let us consider a massless scalar field in the de Sitter background. Due to the curved geometry, the field modes defined in the distant past (\emph{in-modes}) evolve into a linear combination of modes in the distant future (\emph{out-modes}), described by Bogoliubov transformations

\begin{equation}\label{AL}
\hat{a}_{\mathbf{k}}^{\text{in}} = \alpha_k \hat{a}_{\mathbf{k}}^{\text{out}} + \beta_k^* \hat{a}_{-\mathbf{k}}^{\text{out}\dagger},
\end{equation}
where \( \alpha_k \) and \( \beta_k \) are the Bogoliubov coefficients satisfying \( |\alpha_k|^2 - |\beta_k|^2 = 1 \). For the de Sitter case, they are given by

\begin{equation}\label{AM}
|\beta_k|^2 = \frac{1}{e^{2\pi k/H} - 1},
\end{equation}
which corresponds to a thermal spectrum with Gibbons-Hawking temperature \( T = \frac{H}{2\pi} \), where \( H \) is the Hubble parameter. The entangled two-mode squeezed state initially shared by Alice and Bob evolves under the influence of the spacetime background. The reduced state available to Bob becomes mixed due to the coupling with inaccessible field modes. The resulting density matrix for the entangled resource is a thermal-like mixed state

\begin{equation}\label{AN}
\rho_{AB} = \sum_n \lambda_n |n\rangle_A \langle n| \otimes |n\rangle_B \langle n|,
\end{equation}
with occupation probabilities \( \lambda_n = (1 - z) z^n \), where \( z = |\beta_k/\alpha_k|^2 = e^{-2\pi k/H} \). The entanglement is thus degraded with increasing $H$, particularly for low-$k$ modes, reflecting the enhanced particle production induced by the cosmological expansion. Also, the fidelity of teleportation using a thermalized entangled resource is

\begin{equation}\label{AO}
F = \frac{1 + \mathcal{C}}{2},
\end{equation}
where \( \mathcal{C} \) is the concurrence (or alternatively, the entanglement negativity) of the resource state. We emphasize that Eq.~(\ref{AO}) as a well-known result for optimal qubit teleportation using a mixed two-qubit entangled resource under LOCC \cite{Horodecki1999}, is specific to the continuous-variable mode-based teleportation scheme considered here, and should be contrasted with alternative expressions for fidelity in symmetric EPR-like or detector-based teleportation protocols \cite{Koga2018}.
For a two-mode squeezed thermal state, the fidelity is known to be bounded as

\begin{equation}\label{AP}
F = \frac{1}{1 + e^{-2r}},
\end{equation}
where \( r \) is the effective squeezing parameter related to the Bogoliubov coefficients

\begin{equation}\label{AQ}
\tanh r = |\beta_k / \alpha_k|.
\end{equation}
Using this, we obtain

\begin{equation}\label{AR}
F(k) = \frac{1}{1 + e^{-2 \tanh^{-1} (e^{-\pi k/H})}} = \frac{1+e^{-\pi k/H}}{2}.
\end{equation}
This fidelity exhibits a clear and physically transparent behavior. For long-wavelength modes
($k \to 0$), the factor $e^{-\pi k/H}$ approaches unity, and the fidelity tends to
$F_{\rm dS} \to 1$. These modes remain only weakly affected by the expansion, as their
large physical wavelength prevents significant mixing between positive and negative
frequency components. In contrast, short-wavelength modes ($k \to \infty$) yield
$e^{-\pi k/H}\to 0$, leading to the limiting value $F_{\rm dS} \to 1/2$, which represents 
maximal degradation for a two-mode squeezed resource subjected to Gibbons--Hawking 
particle production. The monotonic decrease of $F_{\rm dS}(k)$ with increasing $k$
thus reflects that higher-frequency modes are more efficiently populated through
mode-mixing in de~Sitter spacetime.

\begin{figure}[h]
\centering
\includegraphics[width=0.55\textwidth]{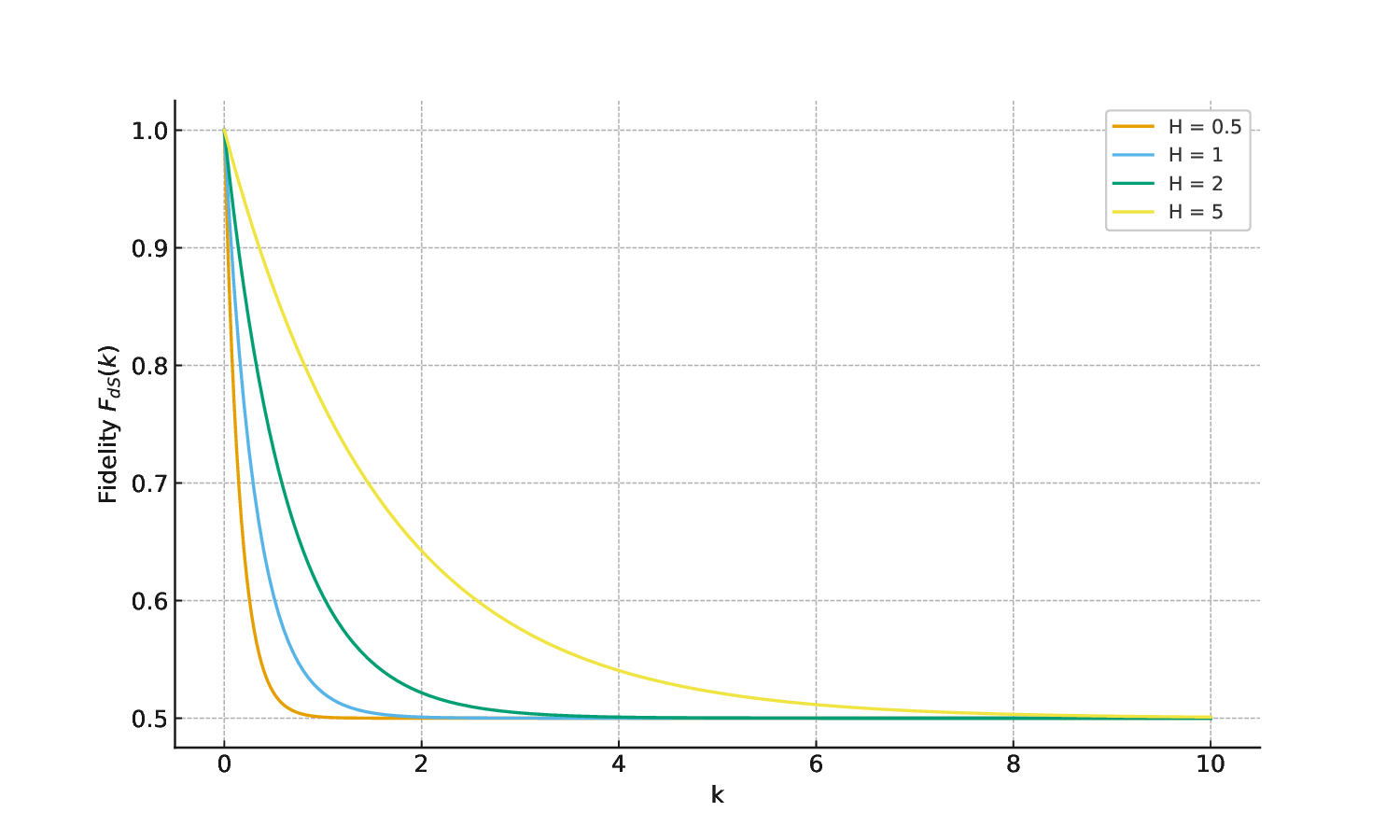}
\caption{
Teleportation fidelity $F_{\text{dS}}(k)$ in de~Sitter spacetime for several values of the Hubble parameter $H$. Larger $H$ increases Gibbons--Hawking particle production, which enhances the Bogoliubov coefficient $\beta_k$ and consequently lowers the fidelity. The limit $H \to 0$ approaches the flat-spacetime baseline.}
\label{fig1}
\end{figure}

To quantify how de~Sitter expansion affects the operational performance of the teleportation protocol, Fig.~\ref{fig1} plots the fidelity $F_{\text{dS}}(k,H)$ as a function of the comoving wavenumber $k$ for several values of the Hubble parameter. Curvature effects are encoded through the Bogoliubov coefficient, whose impact on fidelity depends on both $k$ and $H$. 

For small $H$, the spectrum is sharply peaked near $F \approx 1$, essentially reproducing the flat-spacetime behavior. As $H$ increases, the fidelity decreases uniformly across the entire $k$-range, with the strongest suppression at small $k$ where curvature-induced particle creation is maximal. This behavior directly reflects the enhanced Gibbons--Hawking noise that degrades the shared entangled resource, reducing the effectiveness of continuous-variable teleportation.

\section{Matter-dominated universe}
In a matter-dominated FRW universe, the scale factor in terms of the conformal time is given by

\begin{equation}\label{AS}
a(\eta) = \frac{1}{4} H_0^2 \eta^2,
\end{equation}
where $H_0$ is the present-day Hubble constant. We compute $a''(\eta) = \frac{d^2}{d\eta^2} \left( \frac{1}{4} H_0^2 \eta^2 \right) = \frac{1}{2} H_0^2$, which gives: $\frac{a''(\eta)}{a(\eta)} = \frac{\frac{1}{2} H_0^2}{\frac{1}{4} H_0^2 \eta^2} = \frac{2}{\eta^2}$. Therefore, the mode equation (\ref{K}) becomes

\begin{equation}\label{AT}
\chi_k''(\eta) + \left[k^2 - \frac{2}{\eta^2}\right] \chi_k(\eta) = 0.
\end{equation}
This is the same form as equation (\ref{U}), and the general solution can be expressed in terms of Hankel functions of order $\nu = \frac{3}{2}$:

\begin{equation}\label{AU}
\chi_k(\eta) = \sqrt{|\eta|} \left[ A_k H_{3/2}^{(1)}(k\eta) + B_k H_{3/2}^{(2)}(k\eta) \right],
\end{equation}
where $A_k$ and $B_k$ are constants determined by boundary conditions. To proceed further, one must choose a preferred vacuum (e.g., early-time Minkowski vacuum), and compute the Bogoliubov coefficients $\alpha_k$ and $\beta_k$ by comparing the above solution to the corresponding Minkowski modes in the asymptotic regions.

\begin{figure}[h]
\centering
\includegraphics[width=0.45\textwidth]{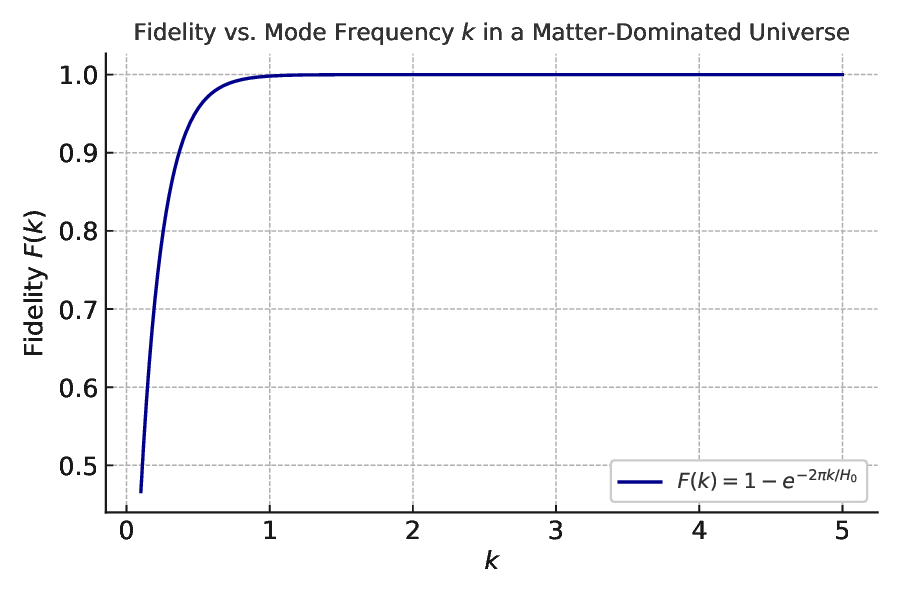}
\caption{Teleportation fidelity as a function of $k$ in a matter-dominated FRW universe. The figure shows that the fidelity gradually increases with $k$ and asymptotically approaches a saturation value.}\label{fig2}
\end{figure}

In the asymptotic past ($\eta \to -\infty$), the space-time is asymptotically flat and the field behaves as in Minkowski space

\begin{equation}\label{AV}
\chi_k^{\text{in}}(\eta) \sim \frac{1}{\sqrt{2k}} e^{-ik\eta}.
\end{equation}
In the far future ($\eta \to 0^-$), the effective potential term $2/\eta^2$ becomes dominant, resulting in particle creation with respect to the Minkowski vacuum. Thus, a Bogoliubov transformation relates the `in' and `out' modes

\begin{equation}\label{AX}
\chi_k^{\text{in}} = \alpha_k \chi_k^{\text{out}} + \beta_k \chi_k^{\text{out}*}.
\end{equation}
A detailed matching procedure yields the Bogoliubov coefficients, and as in previous sections, we adopt a thermal-like approximation for the mean particle number

\begin{equation}\label{AY}
n_k = |\beta_k|^2 \approx \frac{1}{e^{2\pi k/H_0} - 1},
\end{equation}
where $H_0$ is an effective Hubble parameter at late times. This thermal-like approximation allows one to directly identify the mean particle number $n_k$ with the effective noise parameter entering the covariance matrix description of the teleportation channel. In the Gaussian regime, this quantity can be identified with the
smallest symplectic eigenvalue $\nu$ of the partially transposed covariance matrix, which governs the optimal teleportation fidelity \cite{Mari}.
To compute the fidelity, we follow the standard teleportation protocol adapted to curved spacetime, Alice and Bob share an entangled pair of modes initially. Due to the interaction with the curved background, the entangled state evolves into a mixed state from Bob's perspective. It should be emphasized that the global state of the field remains pure under the unitary dynamics generated by the FRW background. However, from Bob’s local perspective, the accessible mode at late times is related to the initial one through a Bogoliubov transformation that entangles it with additional, inaccessible modes. Consequently, after tracing over these degrees of freedom, Bob’s reduced density operator becomes mixed. 
This effective mixedness represents the operational manifestation of entanglement degradation rather than a genuine dynamical loss of purity. As mentioned before, after Alice performs her measurement and communicates the result, Bob applies a local unitary correction to retrieve the teleported state. The output state is then described by a reduced density matrix $\rho_{\text{out}}$, and the fidelity is given by

\begin{equation}\label{AZ}
F = \langle \psi | \rho_{\text{out}} | \psi \rangle = \frac{1}{1 + n_k}.
\end{equation}
Substituting the approximate thermal distribution, the fidelity becomes

\begin{equation}\label{BA}
F(k) = \frac{1}{1 + \frac{1}{e^{2\pi k/H_0} - 1}}=1 - e^{-2\pi k/H_0}.
\end{equation}
This result, as shown in figure \ref{fig2}, confirms that fidelity increases with increasing mode frequency $k$ and is reduced due to particle production effects induced by cosmic expansion. We observe that in the matter-dominated scenario, the fidelity exhibits a relatively slower growth rate compared to the de Sitter universe, but it tends to reach a plateau at late times. This reflects the intermediate expansion behavior of the matter-dominated era, lying between the radiation and accelerated (de Sitter) expansion phases. The behavior of fidelity again confirms the essential role of spacetime curvature and expansion rate in modulating the effectiveness of quantum teleportation protocols in curved spacetimes.

\section{Covariance matrix approach}

The covariance matrix formalism offers an alternative and powerful framework for analyzing Gaussian quantum states, particularly useful when dealing with decoherence and noise induced by external fields or dynamical spacetime backgrounds. In this setup, the effect of cosmological particle creation is effectively modeled as a thermal-like noise channel acting on one or both modes, leading to an effective reduction in entanglement. This formalism provides a complementary and model-independent method for quantifying entanglement degradation in Gaussian states. 
While the teleportation fidelity $F_{\text{FRW}}$ can be directly expressed in terms of the effective squeezing parameter $r_{\text{eff}}$, the covariance matrix allows one to track how all second-order correlations transform under the Bogoliubov mixing of field modes. 
This formalism thus offers a systematic cross-check of the analytical fidelity results obtained in the previous sections and enables a clearer identification of the geometric contribution of spacetime expansion to the overall loss of quantum correlations. 

From section 3, equation (\ref{T}), we can see that the normalized positive-frequency mode function corresponding to the Bunch-Davies vacuum is given by

\begin{equation}\label{BB}
\chi_k(\eta) = \sqrt{-\eta} \, H^{(2)}_\nu(-k\eta),
\end{equation}
where \( H^{(2)}_\nu \) is the Hankel function of the second kind and $\nu = \left| \frac{1 - 3\alpha}{2(1 - \alpha)} \right|$. The original field mode function is

\begin{equation}\label{BC}
\phi_k(\eta) = \frac{1}{a(\eta)} \chi_k(\eta) = \frac{1}{a(\eta)} \sqrt{-\eta} \, H^{(2)}_\nu(-k\eta).
\end{equation}
The covariance matrix of the field modes is constructed from the equal-time two-point functions \cite{weedbrook2012gaussian, AdessoIlluminati2007}

\begin{equation}\label{BD}
\sigma_{ij} = \frac{1}{2} \langle \{ \hat{R}_i, \hat{R}_j \} \rangle,
\end{equation}
where \( \hat{R} = (\hat{q}_k, \hat{p}_k) \) is the phase-space vector for mode \( k \), with

\begin{equation}\label{BE}
\hat{q}_k = \phi_k, \qquad \hat{p}_k = \pi_k = a^2(\eta) \phi_k'.
\end{equation}
Explicitly, the components are

\begin{equation}\label{BF}
\begin{aligned}
\sigma_{qq} &= \langle \hat{q}_k \hat{q}_k \rangle = |\phi_k(\eta)|^2, \\
\sigma_{pp} &= \langle \hat{p}_k \hat{p}_k \rangle = |a^2(\eta) \phi_k'(\eta)|^2, \\
\sigma_{qp} &= \Re\left[ \phi_k(\eta) a^2(\eta) \phi_k'^*(\eta) \right].
\end{aligned}
\end{equation}
These quantities fully characterize the Gaussian state and are the basis for calculating the teleportation fidelity. In the framework of quantum field theory in curved spacetime, the teleportation protocol can be modeled using the vacuum entanglement of a free scalar field. For a pair of modes (say, with wave numbers \( k \) and \( -k \)), the reduced state is Gaussian and fully characterized by its covariance matrix. The fidelity of teleportation of a coherent input state using such an entangled Gaussian channel is given by standard formulae from continuous-variable quantum information theory

\begin{equation}\label{BG}
F= \frac{2}{\sqrt{\det\left(2\sigma + I\right)}},
\end{equation}
where \( \sigma \) is the \( 4 \times 4 \) covariance matrix of the bipartite (two-mode) state, and \( I \) is the identity matrix. In the special case where the two-mode state is symmetric and the input is a vacuum (or coherent) state, the formula simplifies to

\begin{equation}\label{BH}
F= \left[ \det\left( A + \frac{1}{2} I_2 \right) \right]^{-1/2},
\end{equation}
where \( A \) is the local \( 2 \times 2 \) covariance block for one of the modes defined by (\ref{BF}) sufficient to compute the fidelity when the two-mode state
is symmetric. Eqs.~(\ref{BG}) and (\ref{BH}) follow from the standard covariance matrix description of
continuous-variable Gaussian states, under the assumption of symmetric noise
and optimal gain in the teleportation protocol.
In this setting, the teleportation fidelity can be expressed directly in terms
of the effective thermal noise or mean particle number characterizing the state
\cite{weedbrook2012gaussian, Mari}. Alternatively, in terms of the noise number \( \bar{n} \) associated with the teleportation channel we have

\begin{equation}\label{BI}
F= \frac{1}{1 + \bar{n}}.
\end{equation}
Here, $\bar n$ induced by the curved background and plays the same role as $n_k$ in the covariance matrix description,
again corresponding to the symplectic eigenvalue $\nu$. To understand the behavior of the teleportation fidelity in expanding universes, we consider two asymptotic limits of the mode functions:

${\bullet}$ {\it{Subhorizon limit, \( k|\eta| \gg 1 \)}}: In this regime, the Hankel function can be approximated by \cite{Abra}

\begin{equation}\label{BJ}
H_\nu^{(2)}(x) \approx \sqrt{\frac{2}{\pi x}} e^{-i\left(x - \frac{\pi\nu}{2} - \frac{\pi}{4}\right)}, \quad x \gg 1.
\end{equation}
Thus, the mode function becomes

\begin{equation}\label{BK}
\phi_k(\eta) \approx \frac{1}{a(\eta)} \sqrt{-\eta} \, \sqrt{\frac{2}{\pi (-k\eta)}} e^{-i\left(-k\eta - \frac{\pi\nu}{2} - \frac{\pi}{4}\right)} = \frac{1}{a(\eta)} \sqrt{\frac{2}{\pi k}} e^{-i\Theta(k,\eta)},
\end{equation}
with \( \Theta(k,\eta) \) a phase factor. Therefore, the covariance matrix components become approximately

\begin{equation}\label{BL}
\begin{aligned}
\sigma_{qq} &\approx \left| \phi_k(\eta) \right|^2 = \frac{2}{\pi k a^2(\eta)}, \\
\sigma_{pp} &\approx \left| a^2(\eta) \phi_k'(\eta) \right|^2 \approx k \times \text{(constants)}, \\
\sigma_{qp} &\approx 0 \quad \text{(rapid oscillations average to zero)}.
\end{aligned}
\end{equation}
Hence, the fidelity in this regime behaves like

\begin{equation}\label{BM}
F_{\text{sub}} \to \frac{1}{1 + \bar{n}} \approx \text{finite and high},
\end{equation}
which means the teleportation fidelity is close to 1 in subhorizon modes.

${\bullet}$ {\it{ Superhorizon limit, \( k|\eta| \ll 1 \)}}: In this regime, the Hankel function behaves as \cite{Abra}

\begin{equation}\label{BN}
H_\nu^{(2)}(x) \approx \frac{i}{\pi} \Gamma(\nu) \left( \frac{x}{2} \right)^{-\nu}, \quad x \ll 1.
\end{equation}
The mode function becomes

\begin{equation}\label{BO}
\phi_k(\eta) \approx \frac{1}{a(\eta)} \sqrt{-\eta} \, \frac{i}{\pi} \Gamma(\nu) \left( \frac{-k\eta}{2} \right)^{-\nu},
\end{equation}
which gives

\begin{equation}\label{BP}
\begin{aligned}
\sigma_{qq} &\sim |k\eta|^{-2\nu} \\
\sigma_{pp} &\sim |k\eta|^{-2\nu + 2} \\
\sigma_{qp} &\sim \text{nonzero}
\end{aligned}
\end{equation}
The elements of the covariance matrix grow rapidly, and the channel becomes noisy. Hence, the fidelity becomes

\begin{equation}\label{BQ}
F_{\text{super}} \ll 1.
\end{equation}
This signifies that teleportation across superhorizon modes is strongly suppressed due to loss of quantum coherence. For a comprehensive review of the covariance matrix formalism and Gaussian quantum information, see \cite{weedbrook2012gaussian, adesso2014continuous}.

\section{Summary}

In this work, we investigated the performance of quantum teleportation in expanding FRW spacetimes by focusing on how cosmological particle production affects the entanglement structure of quantum fields and, consequently, the teleportation fidelity. Adopting an operationally motivated framework based on a two-mode squeezed vacuum as the entanglement resource, we analyzed the degradation of teleportation fidelity induced by Bogoliubov mode mixing in several cosmologically relevant scenarios.

We considered three representative expansion histories: radiation-dominated, matter-dominated and de Sitter universes. In the radiation-dominated case, conformal invariance of the massless scalar field ensures the absence of particle production, leading to vanishing Bogoliubov coefficients and, consequently, ideal teleportation fidelity identical to that of flat Minkowski spacetime. This result confirms that cosmic expansion alone does not necessarily degrade quantum communication protocols, provided conformal symmetry is preserved.

In contrast, for matter-dominated expansion, particle production becomes nontrivial and induces a gradual reduction of teleportation fidelity. The corresponding Bogoliubov coefficients yield a moderate suppression of fidelity as a function of the comoving momentum, reflecting the relatively mild nature of cosmological particle creation in this background. This behavior is illustrated in Fig. \ref{fig2} and highlights the sensitivity of teleportation performance to the underlying expansion rate.

The most pronounced degradation arises in the de Sitter universe. Owing to the presence of a cosmological horizon and sustained particle production, the teleportation fidelity exhibits a monotonic decrease with increasing comoving momentum, asymptotically approaching the classical limit. As shown in Fig. \ref{fig1}, this strong suppression directly correlates with the Hubble parameter, emphasizing the dominant role of horizon-induced mode mixing in degrading entanglement-based quantum information tasks.

Taken together, these results demonstrate that different cosmological eras imprint distinct signatures on the operational viability of quantum teleportation. While radiation-dominated expansion preserves ideal performance, matter-dominated and de Sitter backgrounds progressively reduce fidelity, with the latter providing the most severe limitation. Our analysis thus clarifies how spacetime dynamics, through particle production and mode mismatch, act as an effective noise source for relativistic quantum communication. Comparison of different cosmological eras and their impact on quantum teleportation studied in this article is summarized in table \ref{tab:cosmo_compare}.

\begin{table}[t]
\centering
\footnotesize
\label{tab:comparison}
\renewcommand{\arraystretch}{1.15}
\begin{tabular}{|c|c|c|p{3.4cm}|p{3.8cm}|}
\hline
\textbf{Cosmological era} 
& \textbf{Scale factor} 
& $\boldsymbol{|\beta_k|^2}$ 
& \textbf{Fidelity} 
& \textbf{Remarks} \\
\hline
Minkowski 
& $a=\text{const}$ 
& $0$ 
& $F = \dfrac{1}{1+e^{-2r}}$ 
& Ideal teleportation, no particle creation \\
\hline
Radiation-dominated 
& $a(t)\propto t^{1/2}$,\; $a(\eta)\propto \eta$ 
& $0$ 
& Same as Minkowski (up to mode mismatch effects)
& Conformal invariance preserves vacuum \\
\hline
Matter-dominated 
& $a(t)\propto t^{2/3}$,\; $a(\eta)\propto \eta^{2}$ 
& $\displaystyle \approx \frac{1}{e^{2\pi k/H_0}-1}$ 
& $F< F_{\text{Mink}}$, increases with $k$ 
& Moderate entanglement degradation \\
\hline
de Sitter 
& $a(t)\propto e^{Ht}$,\; $a(\eta)=-(H\eta)^{-1}$ 
& $\displaystyle \frac{1}{e^{2\pi k/H}-1}$ 
& $F< F_{\text{Mink}}$, decreases with $k$ 
& Maximal particle creation, strongest degradation \\
\hline
\end{tabular}
\caption{Comparison of particle production and teleportation fidelity for different cosmological backgrounds.}
\label{tab:cosmo_compare}
\end{table}

Although our study focuses on an idealized scalar-field model and does not explicitly simulate all steps of the teleportation protocol, the fidelity serves as a meaningful operational indicator of entanglement degradation in curved spacetime. The framework developed here can be extended to include detector-based implementations, massive fields, or more general cosmological scenarios, providing a systematic route toward understanding quantum information processing in dynamical spacetimes.

\end{document}